\documentclass[pdflatex,sn-aps,iicol]{sn-jnl}

\usepackage{graphicx}
\usepackage{bm}
\usepackage{hyperref}
\usepackage{ulem}

\hypersetup{colorlinks,linkcolor=cyan,citecolor=cyan,urlcolor =[rgb]{0.0,0.0,0.5}}
\newcommand{\new}[1]{\textcolor{black}{#1}}

\jyear{2022}%

\providecommand{\myeq}[1]{Eq.~(\ref{#1})}
\providecommand{\mysec}[1]{Sec.~\ref{#1}}
\providecommand{\myfig}[1]{Fig.~\ref{#1}}

\begin{document}

\title[]{Searching Lorentz invariance violation from cosmic photon attenuation}



\author[1]{\fnm{Hao} \sur{Li}}\email{haolee@pku.edu.cn}

\author*[1,2,3]{\fnm{Bo-Qiang} \sur{Ma}}\email{mabq@pku.edu.cn}

\affil[1]{\orgdiv{School of Physics},
  \orgname{Peking University}, \orgaddress{\city{Beijing}~\postcode{100871}, \country{China}}}

\affil[2]{\orgdiv{Center for High Energy Physics}, \orgname{Peking University}, \orgaddress{\city{Beijing}~\postcode{100871}, \country{China}}}

\affil[3]{\orgname{Collaborative Innovation Center of Quantum Matter}, \orgaddress{\city{Beijing}, \country{China}}}

\abstract{Lorentz invariance violation~(LIV) can change the threshold behavior predicted by special relativity and cause threshold anomalies which affect the propagation of cosmic photons. In this work, we focus on the threshold anomaly effect on cosmic photon attenuations by extragalactic background light~(EBL) and discuss how to identify LIV from observations of very high energy~(VHE) photons propagated from long distance in the universe. We point out that the Large High Altitude Air Shower Observatory~(LHAASO), one of the most sensitive gamma-ray detector arrays currently operating at TeV and PeV energies, is an ideal facility for performing such LIV searching. We apply the proposed strategy to discuss the newly observed gamma-ray burst GRB 221009A to demonstrate the predictive ability of our suggestions.}



\subtitle{\footnotesize{Manuscript submitted to EPJC on 13 April 2022, published in Euro.Phys.J.C 83 (2023) 192.
\href {https://doi.org/10.1140/epjc/s10052-023-11334-z}
  {\path{doi:10.1140/epjc/s10052-023-11334-z}}}}

\maketitle

\section{Introduction\label{intro}}

Lorentz invariance, one of the cornerstones of modern physics, plays a fundamental role in constructing theories such as special relativity and quantum field theories which have made great achievements in helping us to understand the nature. As a basic hypothesis, Lorentz invariance has been tested thoroughly and no convincing evidence of its violation exists. However, we still need to consider the possibility of Lorentz invariance violation~(LIV) since experiments cannot rule out extremely small LIV effects. Indeed, certain quantum gravity theories do predict LIV, amongst which are stringy theories~\cite{Amelino-Camelia1997a,Ellis1999,Ellis2000,Ellis2004,Ellis2008,Li2009,Li:2021pre,LI2021104380}, loop quantum gravity~\cite{Gambini1999,Alfaro2002a,Alfaro2002b}, special-relativity-like theories such as doubly special relativity~(DSR)~\cite{AMELINO-CAMELIA2002a,Magueijo2002,AMELINO-CAMELIA2002b, Amelino-Camelia2005,Kowalski-Glikman:2004fsz} and many other theories~\cite{Mattingly2005}. The potential deviation from Lorentz invariance is extremely small since there is still no experiment confirming it and it is believed to be suppressed by the Planck scale $E_\text{Planck}\simeq 1.22\times 10^{19}~\text{GeV}$ for theoretical considerations, e.g., see Ref.~\cite{Amelino-Camelia1998}. As a result, any attempt to uncover the LIV properties of the nature ought to utilize the most sensitive terrestrial or astrophysical experimental approaches, and in the present work we focus on the threshold anomaly method which is based on astrophysical observations of very-high energy~(VHE, $>100~\text{GeV}$) cosmic rays~(especially $\gamma$-rays). According to special relativity~(SR) and quantum electrodynamics~(QED), VHE photons propagating from distant sources can be absorbed by background light in the universe and produce electrons and positrons
~\cite{Nikishov1962,Gould1967a}. Therefore the spectra of cosmic photons may exhibit hard cutoffs and it is more difficult to observe photons with energy beyond thresholds. Nevertheless, LIV can cause threshold anomalies through modified dispersion relations~(MDRs)\footnote{It is also a conceivable speculation that the cross sections obtain modifications as well, but it is model-dependent and the underlying theories are still absent, so we do not consider this effect in this work.}. Because threshold anomalies can change the propagation properties of VHE photons and make the universe more transparent, LIV might be revealed by searching its footprints on astrophysical observations~\cite{KLUZNIAK1999117,Kifune:1999ex,Mattingly2003,Jacobson2003,Li:2021pre2,LI20212254,LI2021a}. In general, both cosmic microwave background~(CMB) and extragalactic background light~(EBL) contribute to the attenuations, and in this work we focus on EBL which is more sensitive to TeV photons while CMB is more sensitive to PeV photons~\cite{Ruffini2016}. For this purpose, The Large High Altitude Air Shower Observatory~(LHAASO) provides a unique opportunity to carry out this kind of researches for its ability to explore the gamma-ray sky above TeV\@. Future observations of LHAASO of VHE photons from distant objects such as gamma-ray bursts~(GRBs) and active galactic nuclei~(AGNs) can help us acquire a better understanding of LIV, therefore we present an analysis of the potential indications of the future results from LHAASO on LIV by combining threshold anomalies and EBL\@.

This paper is organized as follows. In \mysec{livthres} we provide a brief introduction to LIV models, MDRs~(\mysec{livmodel}) and LIV induced threshold anomalies~(\mysec{thresanomaly}). We then give a brief review about VHE photon attenuations by EBL and some conceivable results on which we base to discuss future LHAASO observations and LIV in \mysec{ebl}. After introducing these basic materials, we analyze possible footprints of LIV induced threshold anomalies on results of LHAASO in \mysec{discussion}, and thus future data of LHAASO can be used to constrain LIV threshold anomalies and some LIV parameters.

\section{Lorentz invariance violation, modified dispersion relations and threshold anomalies\label{livthres}}

In this section, we present a brief introduction to the theoretical setups of LIV, LIV induced (low-energy effective) modified dispersion relations~(MDRs) and LIV induced threshold anomalies, and more detailed descriptions can be found in reviews like~\cite{Mattingly2005,Jacobson2006,Liberati2009,Amelino-Camelia2013} and references therein.

\subsection{Lorentz invariance violation and modified dispersion relations\label{livmodel}}

LIV emerges from many approaches to QG\@. Almost all models of LIV suggest modified dispersion relations~(MDRs), which prove to be crucial for LIV phenomenological study~\cite{Mattingly2005, Amelino-Camelia2013}. However, no matter what the true fundamental theories are, we can always adopt a model-independent formulation~\cite{Xiao2009,Shao2010b} of the MDR of photons of which the energy is well below $E_\text{Planck}$\footnote{This condition is satisfied for photons that can be observed by existing observatories.}:

\begin{equation}
  \label{generalmdr}
  \omega^2(k)=k^2+\frac{s}{E_\text{LV}}k^3 +O\left(k^4{(1/E_\text{Planck})}^2\right),
\end{equation}
where $\omega$ and $k$ are the energy and the magnitude of the momentum of the photon respectively, $s=\pm 1$ which corresponds to superluminal and subluminal cases respectively or $s=0$ which means there is no linear correction to the speed of light\footnote{The speed of light is assumed to be defined as $v=\partial\omega(k)/\partial k$.} for some theoretical considerations, and $E_\text{LV}$ is a suppression parameter of the order of magnitude of $E_\text{Planck}$. Since higher orders of \myeq{generalmdr} are more severely suppressed by the Planck scale, we could drop them safely as long as we only deal with TeV or PeV photons and therefore we obtain a simplified formulation\footnote{We do not discuss those models without the first-order correction, i.e., $s=0$, because most theories and some phenomenological researches favour a linear correction to the speed of light.}:

\begin{equation}
  \omega^2(k)=k^2-\xi k^3,\label{mdr}
\end{equation}
where we further define a new parameter $\xi=-s/E_\text{LV}$ for convenience. This expression is what we adopt to discuss the LIV induced threshold anomalies.

Before going on, we would like to briefly discuss the sign in \myeq{generalmdr} or \myeq{mdr} which means possible helicity-dependence of the speed of light. We can calculate the speed of light according to \myeq{generalmdr},

\begin{align}
  v(E) & \equiv\frac{\partial\omega(k)}{\partial k}\notag                                      \\
       & =c\left(1\pm\frac{E}{E_\text{LV}}\right) = c\left(1-\xi E\right),\label{speedoflight}
\end{align} %
which exhibits both superluminal~(``$+$'') and subluminal~(``$-$'') features. If both the signs in \myeq{speedoflight} can be taken, then those theories predicting this feature are helicity-dependent. However, phenomenological explorations do not favour helicity-dependent linear corrections to the speed of light~(see, {\it e.g.\/} Refs.~\cite{Shao2011,wei2021tests} and references therein). Therefore we should either believe that the leading corrections are at least second order or look for models which only predict one of these two signs. In fact, the string-inspired model~\cite{Ellis2000,Ellis2008,Li2009,Li:2021pre,LI2021104380} convince us that there could be a helicity-independent correction to the speed of light which is always subluminal, and meanwhile some researches using the data of GRBs~\cite{Shao2010f, Xiao2009a, Zhang2015, Xu2016a, Xu2016,  Xu2018, Liu2018, ZHU2021136518, Chen2021}
and AGNs~\cite{Li2020} suggest that this is conceivable. As we will see, it is the subluminal MDR that cause the most interesting threshold anomalies.

\subsection{Threshold anomalies\label{thresanomaly}}

There is an urgent problem with which the developments of quantum gravity~(QG) are confronted: experimental evidence or guidance is in badly need. LIV phenomenology is thus the key to our understanding of the true underlying theories. In this work we concentrate on the research on the threshold anomaly approach to LIV phenomenological study. A quick review of LIV induced threshold anomalies and the resulting observable phenomena are given hereafter, and details can be found in Refs.~\cite{Mattingly2003, Jacobson2003,Li:2021pre2,LI20212254,LI2021a} {\it etc\/}.

Let us consider the process $\gamma\gamma\to e^-e^+$, which contributes mainly to the attenuations of VHE photons with energy less than several PeVs\footnote{For (above) PeV photons, it is the process $\gamma\gamma\to e^-e^+e^-e^+$ that contributes mostly~\cite{Ruffini2016}, while we do not pay attention to this situation.}. According to special relativity~(SR), there is a lower threshold\footnote{A threshold means that when we fix the energy of one photon, the smallest~(lower threshold) or the largest~(upper threshold) energy of the other photon that makes the process occur kinematically.} which can be calculated as follows. Since in SR Lorentz invariance is still preserved, we conclude that there is a lower threshold and the threshold occurs when the two ingoing photons are head-to-head which is obvious if one considers this process in the center-of-mass reference frame and then transforms it into the experimental one by proper Lorentz transformations. As a result, we can take the threshold configuration:

\begin{align}
  p_1     & =(E,0,0,E),\notag                                                                    \\
  p_2     & =(\varepsilon_b,0,0,-\varepsilon_b),\notag                                           \\
  p_3=p_4 & =\left(\frac{E+\varepsilon_b}{2},0,0,\frac{E-\varepsilon_b}{2}\right),\label{config}
\end{align}
where $p_1$ and $p_2$ are the 4-momenta of the two ingoing photons, $p_3$ and $p_4$ are the 4-momenta of the outgoing electron and positron, and we always assume that the photon with energy $\varepsilon_b$ is from background light which meanwhile means that $\varepsilon_b\ll E$. Then the mass-shell condition reads

\begin{equation}
  m_e^2\equiv p_3^2 = {\left(\frac{E+\varepsilon_b}{2}\right)}^2-{\left(\frac{E-\varepsilon_b}{2}\right)}^2,\label{massshell}
\end{equation}
which immediately yields the threshold condition for $\gamma\gamma\to e^-e^+$:

\begin{equation}
  E\ge E_\text{th}=\frac{m_e^2}{\varepsilon_b}.\label{threshold}
\end{equation}
If the energy of a photon exceeds $E_\text{th}$, then the photon can be absorbed with certain configurations. The footprints of this process can be thus distinguished by the suppression of the spectrum above this threshold. Otherwise, the photon can propagate freely if its energy is less than $E_\text{th}$ since this process is kinematically forbidden. To be more specific, we calculate typical thresholds for CMB and EBL according to \myeq{threshold} for future reference. For CMB, we take its mean energy which is $\varepsilon_b\simeq 6.35\times 10^{-4}~\text{eV}$, and therefore $E_\text{th}^\text{CMB}\simeq 411~\text{TeV}$; for EBL we estimate the energy $\varepsilon_b$ according to \myfig{ebl_example}, and we take $\varepsilon_b\simeq 10^{-3}~\text{eV}$ to $1~\text{eV}$ which gives $E_\text{th}^\text{EBL}\simeq 261~\text{GeV}$ to $261~\text{TeV}$. As we can see, photons with energy less than $E_\text{th}^\text{CMB}$ could be absorbed by EBL and about 260 GeV is enough for a photon to be absorbed by EBL\@. Although $E_\text{th}^\text{EBL}$ is a fairly rough estimation since unlike CMB, the distribution of EBL is more complicated, it is sufficient for qualitative discussions and analyses.

However, once LIV~({\it i.e.\/}~\myeq{mdr}) is taken into consideration, the threshold property of $\gamma\gamma\to e^-e^+$ is altered radically, especially for subluminal photons~($\xi>0$). Detailed analyses are present in Refs.~\cite{Mattingly2003,Jacobson2003,LI2021a} already, and we list the main results as follows.
We adopt the same configuration as in \myeq{config} but with $p_1=(E,0,0,E)$ replaced by $p_1=(\omega(k),0,0,k)$ where $\omega(k)$ is given by \myeq{mdr}\footnote{We do not change $p_2$ because we have already assumed that $\varepsilon_b$ is so small that its LIV effect is negligible.}. We also assume that the dispersion relation of electron or positron is preserved for the existence of both theoretical suggestions~\cite{Ellis2008} and experimental evidence~\cite{Maccione2007,Ellis2009, Li2022}. There is another obstacle to analyzing threshold anomalies arising from LIV\@: the energy-momentum conservation law may be modified. For example, in DSR it is the ordinary Lorentz invariance/covariance that is violated while the novel form of the Lorentz invariance/covariance is realized by modified (nonlinear) group generators. Therefore the energy-momentum conservation law in DSR should be covariant under the what is called deformed Lorentz symmetry and the simple additive formulae are invalid in general~\cite{AMELINO-CAMELIA2002a,AMELINO-CAMELIA2002b,Judes2003}. But since DSR is just one of the efforts to understand QG, the correct energy-momentum conservation law can certainly be very different from that of DSR\@. However, we have no aspiration to judge which model is more credible, thus we only make a reasonable assumption that at least in some reference frames, the ordinary additive formulae are still valid and we only focus on discussing threshold anomalies in these reference frames. Then the mass-shell condition for electrons reads~(for $k>0$):

\begin{equation}
  m_e^2=\frac{-\xi k^3 + 4\varepsilon_b k}{4}+O(\xi^2)+O(\xi\varepsilon_b).
  \label{massshell2}
\end{equation}
Based on the threshold condition in \myeq{massshell2}, a recent analysis~\cite{LI2021a} proposes an interpretation to the threshold behaviors which fall into three different cases according to the different values of $\xi$:

\begin{description}
  \item[Case I] If $\xi>\xi_c={16\varepsilon_b^3}/{(27m_e^4)}$, subluminal photons cannot be absorbed by background photons with energy $\varepsilon_b$ through the process $\gamma\gamma\to e^- e^+$.
  \item[Case II] If $0<\xi<\xi_c$, a subluminal photon can be absorbed only when its energy falls into a certain closed interval with its lower bound greater than $E_\text{th}$. This means that there is an upper threshold and the $\varepsilon_b$ background is again transparent to photons with energy exceeding this upper threshold.
  \item[Case III] If $\xi<0$, the threshold behavior resembles that of the case in SR, except that the threshold now is smaller than $E_\text{th}$.
\end{description}
It should be noted that the aforementioned interpretation is only valid for photons with energy not too large, and it is enough for our discussion in this work because this interpretation is reliable for EeV photons while we only consider photons with energy less than several hundred TeVs. It is also noteworthy that {\it Case I} and {\it Case II} are quite intriguing. In both of these cases, the $\varepsilon_b$ background can be transparent to VHE photons which disagrees with the case in SR where photons can be absorbed once their energy is larger than $E_\text{th}$.

\section{VHE photon attenuations by EBL\label{ebl}}

In this section, we give a brief review about VHE photon attenuations by EBL and a plausible EBL model~\cite{Dominguez2011} as well as its results is adopted. These materials as well as those in \mysec{livthres} then can serve as the starting point for discussing potential footprints of LIV on future observations of LHAASO\@.

\begin{figure}
  \centering
  \includegraphics[scale=0.85]{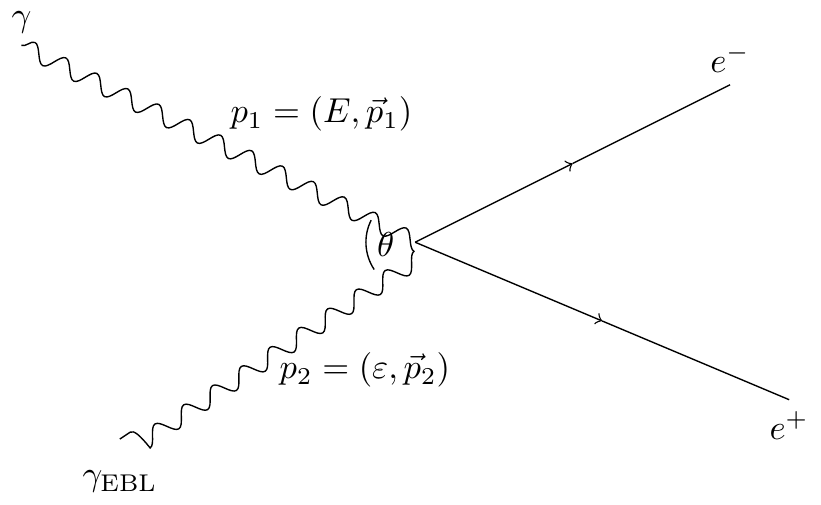}
  \caption{\label{geom}The diagram of the pair production process: $\gamma\gamma\to e^-e^+$.}
\end{figure}

For VHE photons, two major processes contribute to their attenuations. One is the pair production process $\gamma\gamma\to e^- e^+$~\cite{Nikishov1962, Gould1967a}. Another one is the double pair production process $\gamma\gamma\to e^- e^+ e^- e^+$~\cite{Ruffini2016}. It is shown that for TeV photons and attenuations by EBL, we can only concentrate on the first one, {\it i.e.\/}, pair production process~\cite{Ruffini2016}. We consider the process shown in \myfig{geom}, where a VHE photon with energy $E$ interacts with a EBL photon with energy $\varepsilon$ and produces an electron and a positron. The angle between these two photons is $\theta$.

The cross section of this process can be calculated~\cite{Breit1934,Gould1967b}, which is

\begin{align}
   & \sigma_{\gamma\gamma}(E,\varepsilon,\theta) = \notag                                                                            \\
   & \ \ \ \ \ \ \ \ \ \frac{3\sigma_T}{16}\left[2\beta(\beta^2-2)+(3-\beta^4)\ln\frac{1+\beta}{1-\beta}\right],\label{crosssection} \\
   & \beta =\sqrt{1-\frac{2m_e^2 c^4}{E\varepsilon (1-\cos\theta)}},\label{crosssection1}
\end{align}
where $\sigma_T$ is the Thompson scattering cross-section and we restore the constants $\hbar$ and $c$. One can tell the expected threshold condition in SR immediately from \myeq{crosssection1} since it requires $E\ge 2m_e^2 c^4/\left(\varepsilon(1-\cos\theta)\right)$ and if we take $\theta=\pi$ we obtain the formulation of the threshold condition in \myeq{threshold}. With this cross section we can define the optical depth $\tau$~\cite{Dominguez2011,Gilmore2012}:

\begin{align}
   & \tau (E_\text{obs},z) = \int^z_0 \frac{dl}{dt}dt\int^1_{-1}d\mu (1-\frac{\mu}{2})\notag                                                                                                                \\
   & \ \ \ \ \ \ \ \ \ \ \ \ \,\times\int^\infty_{\varepsilon_\text{min}}d\varepsilon^\prime \sigma_{\gamma\gamma}(E_\text{obs}(1+t),\varepsilon^\prime,\theta)n(\varepsilon^\prime,t),\label{opticaldepth} \\
   & \varepsilon_\text{min}=\frac{2m_e^2 c^4}{E_\text{obs}(1+t)\mu},                                                                                                                                        
\end{align}
where $\mu$ stands for $\left(1-\cos\theta\right)$, $n$ is the local EBL photon number density, $E_\text{obs}$ is the observed energy of the photon, $z$ is the redshift of the source, and

$$
  \frac{dl}{dt}=\frac{c}{H_0(1+t)\sqrt{\Omega_m(1+t){}^3+\Omega_\Lambda}}
$$
with the $\Lambda$CDM model parameters $H_0$, $\Omega_m$ and $\Omega_\Lambda$ which describes the cosmological effects on the propagation of photons.
The role played by the optical depth can be clearly understood via considering the relation between the observed and intrinsic flux:

\begin{equation}
  F_\text{obs}=F_\text{int}\times e^{-\tau(E,z)},\label{flux}
\end{equation}
which means the larger the optical depth is, the harder for the flux to be observed.

\begin{figure*}
  \includegraphics[scale=0.8]{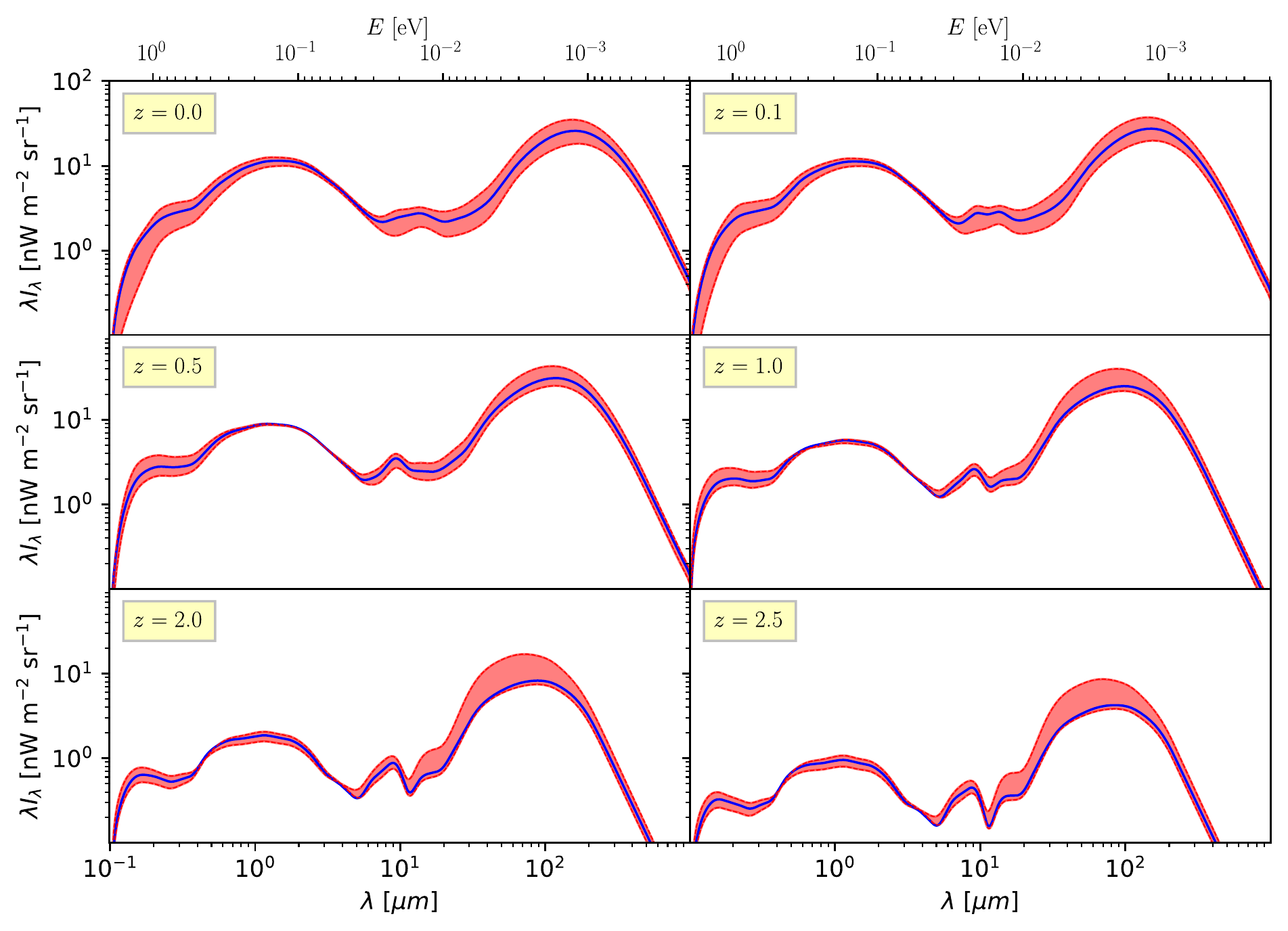}
  \caption{\label{ebl_example}EBL in a comoving frame predicted by the model of Dom{\'\i}nguez {\it et al.} The distributions of EBL at redshift $z=0, 0.1, 0.5,1.0,2.0$ and $2.5$ are shown and more details can be found at~\url{http://side.iaa.es/EBL/} and in Ref.~\cite{Dominguez2011}. The blue solid lines represent the distributions of EBL, and the red dashed lines as well as the red shadowed regions show the uncertainties. The horizontal axes are shown in both the units $\mu\text{m}$~(lower axes) and $\text{eV}$~(upper axes).}
\end{figure*}

\begin{figure*}
  \includegraphics[scale=0.9]{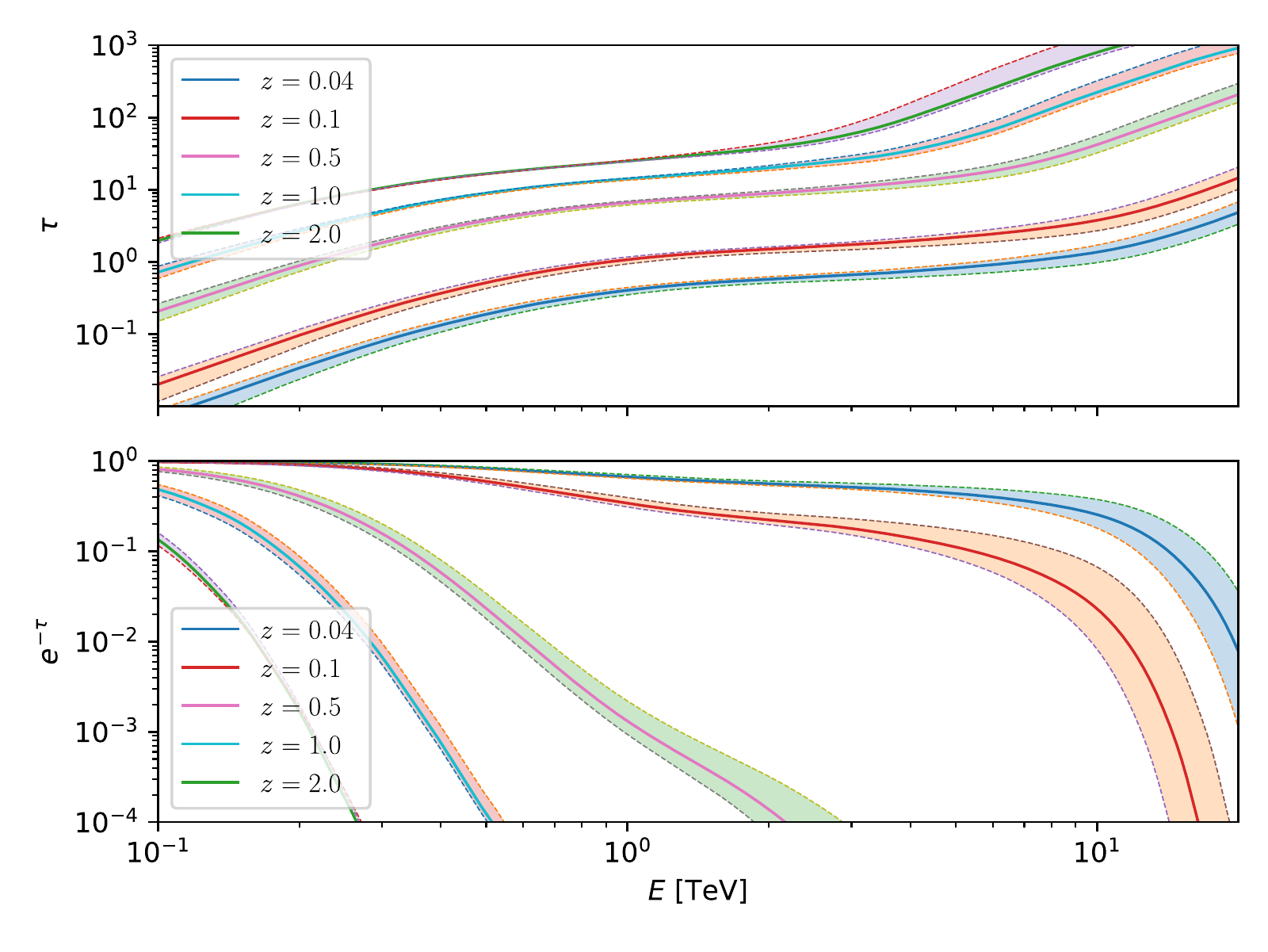}
  \caption{\label{tau}The optical depth~(upper panel) and the exponential of the minus optical depth~(lower panel) versus energy predicted by the model of A. Dom{\'\i}nguez {\it et al.}~\cite{Dominguez2011} for different redshifts $z=0.04,0.1,0.5,1.0$ and $2.0$. More data and details are publicly available at~\url{http://side.iaa.es/EBL/}. The dashed lines and shadowed regions show the uncertainties.}
\end{figure*}

In order to determine the optical depth, we need to first know the distribution of EBL photons in the universe. The most reliable way is to detect the EBL density directly but it is quite difficult because of significant backgrounds. As a result, utilizing both \myeq{opticaldepth} and observations of AGNs, GRBs and other astrophysical objects to constrain $n$ indirectly becomes one of the most important approaches to determine it. Modeling or constraining EBL is a large and active filed with lots of problems to be solved, and it is beyond the scope of this article to discuss its details. In this article we only use one of those plausible models of EBL which is given by Dom{\'\i}nguez {\it et al.}~\cite{Dominguez2011}, and all the data used in the following can be found at~\url{http://side.iaa.es/EBL/}. Several typical distributions of EBL with different redshifts predicted by Dom{\'\i}nguez {\it et al.} are exhibited in \myfig{ebl_example}. With the model of EBL distributions, the optical depth can be calculated straightforwardly by utilizing \myeq{opticaldepth}. For example, \myfig{tau} shows some typical results of the optical depth and the exponential of minus optical depth at different redshifts, and we can conclude from it that the flux of a source declines as its redshift or the energy increases according to \myeq{flux}. We then perform detailed analyses based on the results of $\tau$ and $e^{-\tau}$ with specific redshifts in the next section.


\section{Discussions\label{discussion}}

\subsection{Basic information about LHAASO\label{basiclhaaso}}

The Large High Altitude Air Shower Observatory or LHAASO for short is a new generation detector array that is capable of revealing the very high energy universe. It locates at high altitude~(about 4400 m) in Daocheng, Sichuan Province, China, with high sensitivity to gamma-rays from about 20~GeV up to PeV. It is composed of KM2A~(1 $\text{km}^2$ array fro electromagnetic particle detectors, or ED and also muon detectors, or MD), WCDA~(water Cherenkov detector) and WFCTA~(wide field-of-view air Cherenkov telescopes). The results of LHAASO provide us an opportunity to search LIV in the TeV to PeV energy band where previous experiments cannot. Indeed, it is quite encouraging that the LHAASO collaboration discovers a dozen of photons with energy about and exceeding $1~\text{PeV}$ only by analyzing the data of the first year~\cite{Cao:2021}. As a result, it is promising that LHAASO is able to shed light on the studies of LIV with more VHE or even ultra-high energy~(UHE) photons observed. We thus think it is necessary to discuss how to identify LIV from observations of LHAASO in the following from the point of view of VHE photon attenuations by EBL and LIV induced threshold anomalies. More details of LHAASO can be found in Refs.~\cite{Cao2010,DISCIASCIO2016166, Cao2019, Bai:2019khm}.

\subsection{The strategy for correlating observations of LHAASO to potential sources\label{correlation}}

The LHAASO collaboration has its own strategies for correlating gamma-ray observations to potential sources, however, these strategies depend on standard descriptions of the universe and particle physics which are of course without LIV\@. In order to study LIV from the data of LHAASO, we need a new strategy which is able to contain LIV effects, especially for temporal correlations. For this purpose, based on the criteria described in Refs.~\cite{Amelino-Camelia2016, Amelino-Camelia2017, Huang2018}, we present strategies comprised of the temporal correlation strategy and the angular correlation strategies in the following.

The temporal correlation strategy is based on the time of light flight approach to LIV phenomenology. According to Ref.~\cite{Jacob2008}, the LIV induced time of light flight difference arising from \myeq{mdr} for two photons with energy $E_h$ and $E_l$ respectively emitted from a source simultaneously is

\begin{equation}
  \Delta t_\text{LV}= \xi\frac{E_h-E_l}{H_0}\int_0^z \frac{(1+z^\prime)dz^\prime}{\sqrt{\Omega_m(1+z^\prime){}^3+\Omega_\Lambda}}, \label{tlv1}
\end{equation}
where $z$ is the redshift of the source, $[\Omega_m, \Omega_\Lambda]=[0.315^{+0.016}_{-0.017}, 0.685^{+0.017}_{-0.016}]$ are parameters of the FRW cosmology and $H_0=67.3\pm 1.2~ \text{km}~\text{s}^{-1}\text{Mpc}^{-1}$ is the present-day Hubble constant~\cite{2014cgo}. We omit $E_l$ because most cases in studies satisfy $E_h\gg E_l$ and render \myeq{tlv1} to be

\begin{equation}
  \Delta t_\text{LV} = (1+z) \xi K, \label{tlv2}
\end{equation}
where we define

$$
  K = \frac{E_h}{H_0(1+z)}\int_0^z \frac{(1+z^\prime)dz^\prime}{\sqrt{\Omega_m(1+z^\prime){}^3+\Omega_\Lambda}}.
$$
The observed time difference $\Delta t_\text{obs}$ originate from the combination of both \myeq{tlv2} and the intrinsic time difference $\Delta t_\text{int}$, therefore we have

\begin{equation}
  \Delta t_\text{obs} = \Delta t_\text{LV} + (1+z)\Delta t_\text{int}. \label{tlv3}
\end{equation}
Then we can check the temporal correlation between the photon and the source~(GRB or AGN) utilizing the following criteria~\cite{Amelino-Camelia2016, Amelino-Camelia2017, Huang2018}. From Eq.~(\ref{tlv3}) we have

\begin{equation}
  \left\lvert \frac{\Delta t_\text{obs}}{1+z}-\Delta t_\text{int}\right\rvert=\left\lvert \xi K\right\rvert, \label{timecrit}
\end{equation}
and then we can take $\xi K$ as a time window where we search the possible sources.
For example, we assume $z=1$ and $E_h=15 ~\text{TeV}$, and we take an optional constraint of $\xi$ which is $\xi^{-1}\geq 3.6\times 10^{17}~\text{GeV}$~\cite{Shao2010f, Xiao2009, Xiao2009a, Shao2010b, Zhang2015, Xu2016a, Xu2016,  Xu2018, Liu2018, ZHU2021136518, Li2020}. Then we have $\left\lvert \xi K \right\rvert \leq 3~\text{hours}$. Therefore when we want to search the possible sources, sources that are observed a couple of hours earlier or later can be taken as candidates.

Next we focus on the criteria for direction or angular correlations. The strategies to establish correlations between VHE photons and sources depend on the form of the observed data. There are separated photons and diffused photons for which we should adopt very different methods to establish correlations. For separated photons, we can utilize the following Gaussian to check the correlations~\cite{Amelino-Camelia2016, Amelino-Camelia2017, Huang2018}:

\begin{equation}
  P(\gamma, \text{GRB})=\frac{1}{\sqrt{2\pi \sigma^2}}\exp(-\frac{\Delta\Phi^2}{2\sigma^2}),\label{directioncriterion}
\end{equation}
where $\Delta\Phi $ is the angular separation between the source and the photon, and $\sigma=\sqrt{\sigma^2_\text{source}+\sigma^2_\gamma}$ is the standard deviation of the angular uncertainties of the positions of the source and the photon. A source and a photon are correlated if the angular separation between them are less than $3\sigma{}$~\cite{Huang2018}. While for diffused photons, the situation is more complicated. Because of the extremely high-level background, the distributions of diffused photons are almost uniform. In order to recognize potential excess of photons around a source, we utilize an ``angular averaged'' criterion. Specifically, we first determine a source~(GRB or AGN), then we count the number of photons $N_\gamma\left(\Omega\right)$ within a specific energy band in a circular area around the source with the (spherical) angle $\Omega$ and define the angular averaged number density $\rho\left(\Omega\right)=N_\gamma/\Omega$. If there is no excess, $\rho\left(\Omega\right)$ will be almost a constant as we decrease $\Omega$. However, if $\rho\left(\Omega\right)$ increases significantly as we decrease $\Omega$, we can establish a correlation between the photons and the source spatially. However, the possible excess could be not significant enough, we may use the following method to solve this problem. Again we consider a Guassian function

\begin{equation}
  f(\rho; \Sigma,\rho_c)=\frac{1}{\sqrt{2\pi\Sigma^2}}\exp\left(-\frac{(\rho-\rho_c){}^2}{2\Sigma^2}\right), \label{gauss2}
\end{equation}
where $\Sigma$ and $\rho_c$ are free parameters to be determined with the method given in the following and $\rho=\rho(\Omega)$. There is a famous relation which inspires us:

$$
  \lim_{\Sigma \to 0}f(\rho;\Sigma,\rho_c)=\delta (\rho-\rho_c).
$$
Indeed, the peak centered at $\rho_c$ of this Guassian gets sharper as $\Sigma$ becomes smaller. If we consider the ratio

\begin{align}
  r & \equiv 1-f(\rho;\Sigma,\rho_c)/f(\rho_c;\Sigma,\rho_c)\notag \\
    & =1-\exp(-\frac{(\rho-\rho_c){}^2}{2\Sigma^2}),\notag
\end{align}
which represents the relative change of this Guassian, we can conclude that for the same amount of the deviation $\left\vert\rho-\rho_c\right\vert$, $r$ is bigger for a smaller $\Sigma$. Therefore, so long as we take a very small value of $\Sigma$, we are able to reveal tiny excesses of diffused photons. However we cannot choose an arbitrarily small $\Sigma$, an improperly small $\Sigma$ may make us misidentify fluctuations as excesses and lead to wrong correlations. As a result, we determine the parameters in \myeq{gauss2} as follows. Since $\rho_c$ is the angular averaged number density of backgrounds, we can choose a large enough spherical angle $\Omega_0$ and determine $\rho(\Omega_0)$, and conclude approximately that $\rho_c\simeq\rho(\Omega_0)$. In order to avoid the influence of fluctuations, we can choose different regions both regular and even irregular to determine $\rho(\Omega_0)$ and take the averaged value as a good approximation of $\rho_c$. After finishing the determination of $\rho_c$, we can further determine $\Sigma$. Now that we obtain the values of $\rho$ for different regions, each of them can be considered as $\rho_c$ plus the fluctuation, then we look for the smallest value of $\Sigma$ that makes $r\ge 0.9$ for example\footnote{We can choose this boundary to make the criterion more sensitive~(larger $r$) or insensitive~(smaller $r$).} and the corresponding $\Sigma$ is just what we need.

\subsection{LIV from future observations of LHAASO\label{livfromlhaaso}}

\begin{figure}
  \includegraphics[scale=0.51]{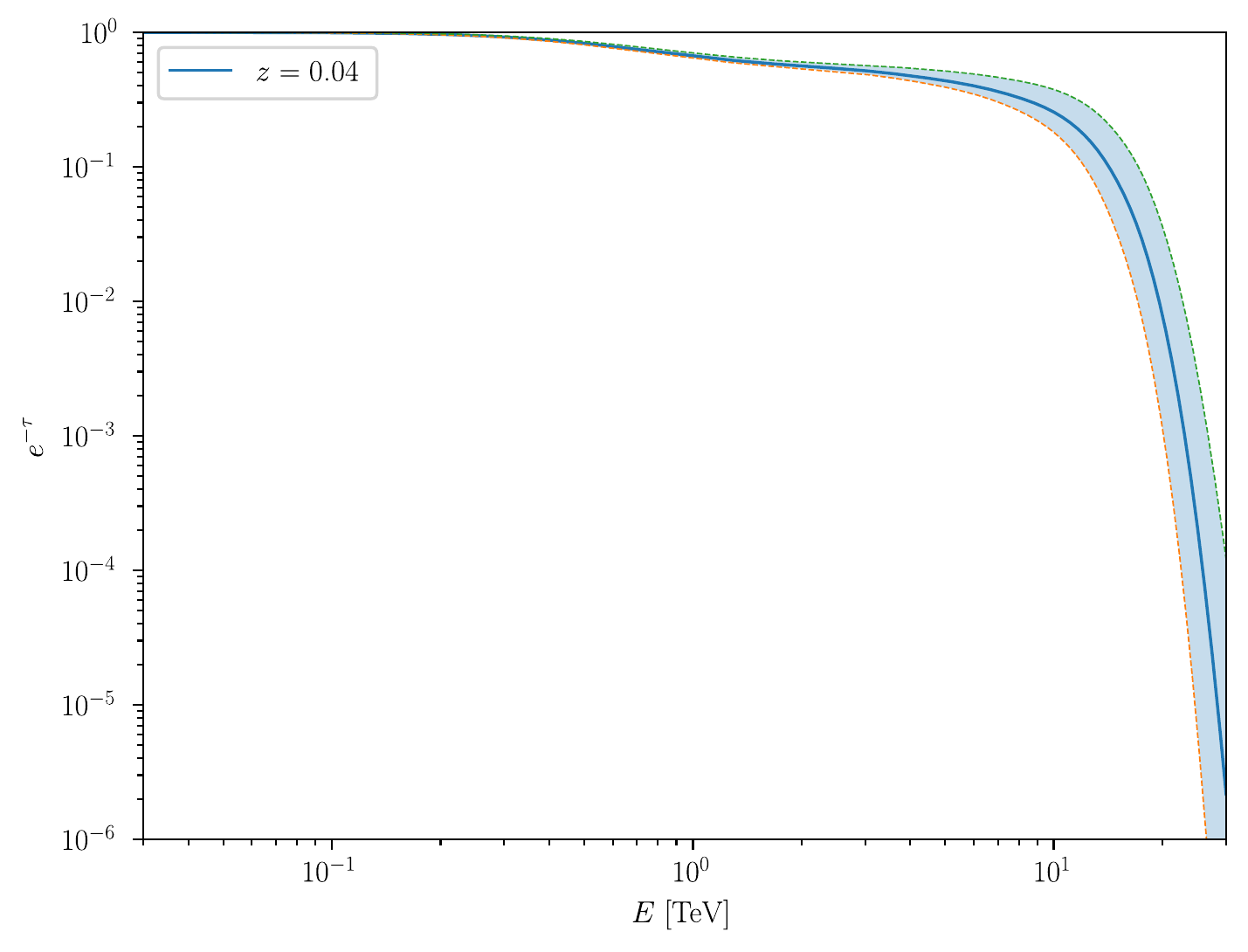}
  \caption{\label{tau0.04}The EBL attenuation~($e^{-\tau}$) of photons from a source with redshift $z=0.04$. The model and data are taken from Ref.~\cite{Dominguez2011} and \url{http://side.iaa.es/EBL/}.}
\end{figure}

\begin{figure}
  \includegraphics[scale=0.51]{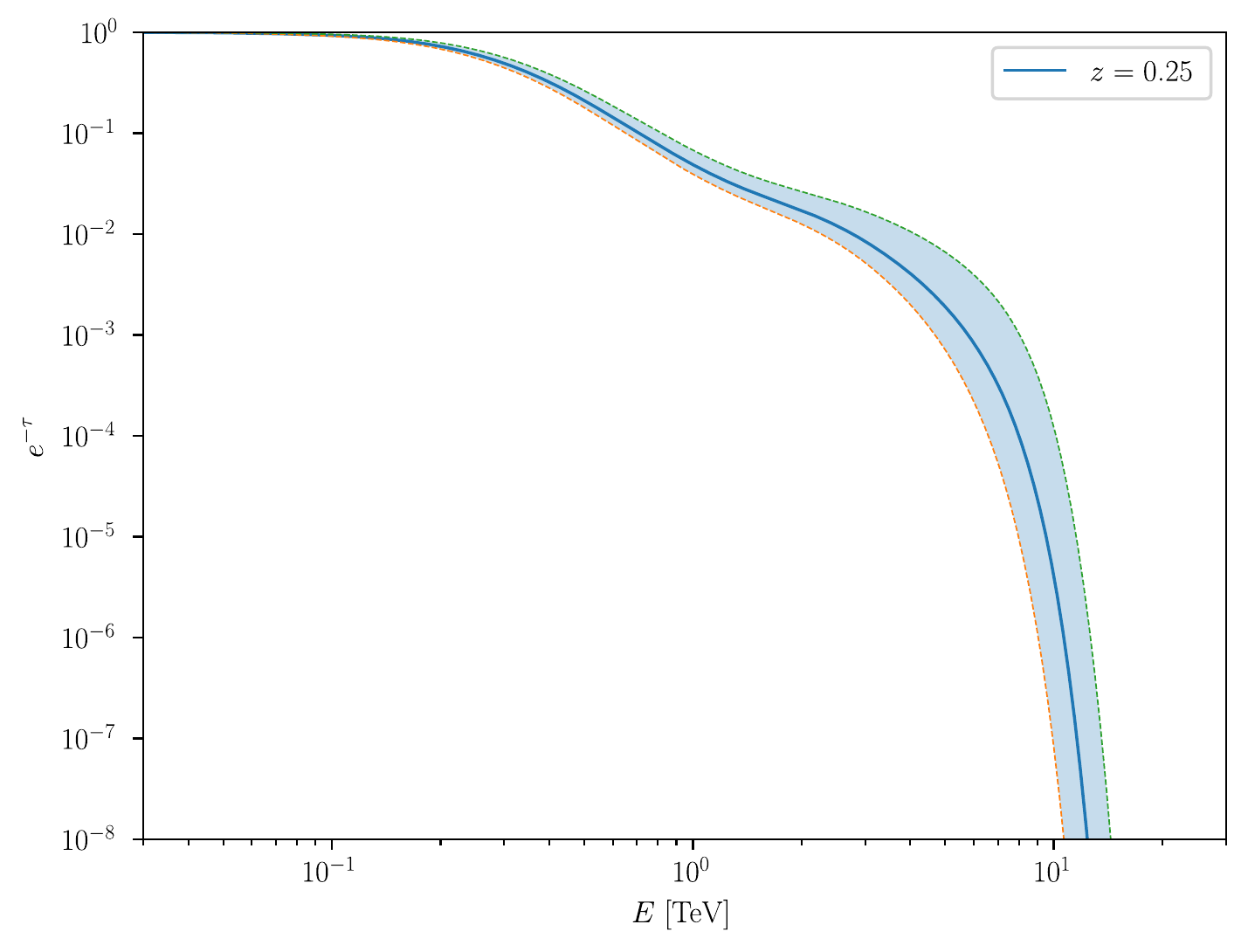}
  \caption{\label{tau0.25}The EBL attenuation~($e^{-\tau}$) of photons from a source with redshift $z=0.25$. The model and data are taken from Ref.~\cite{Dominguez2011} and \url{http://side.iaa.es/EBL/}.}
\end{figure}

\begin{figure}
  \includegraphics[scale=0.51]{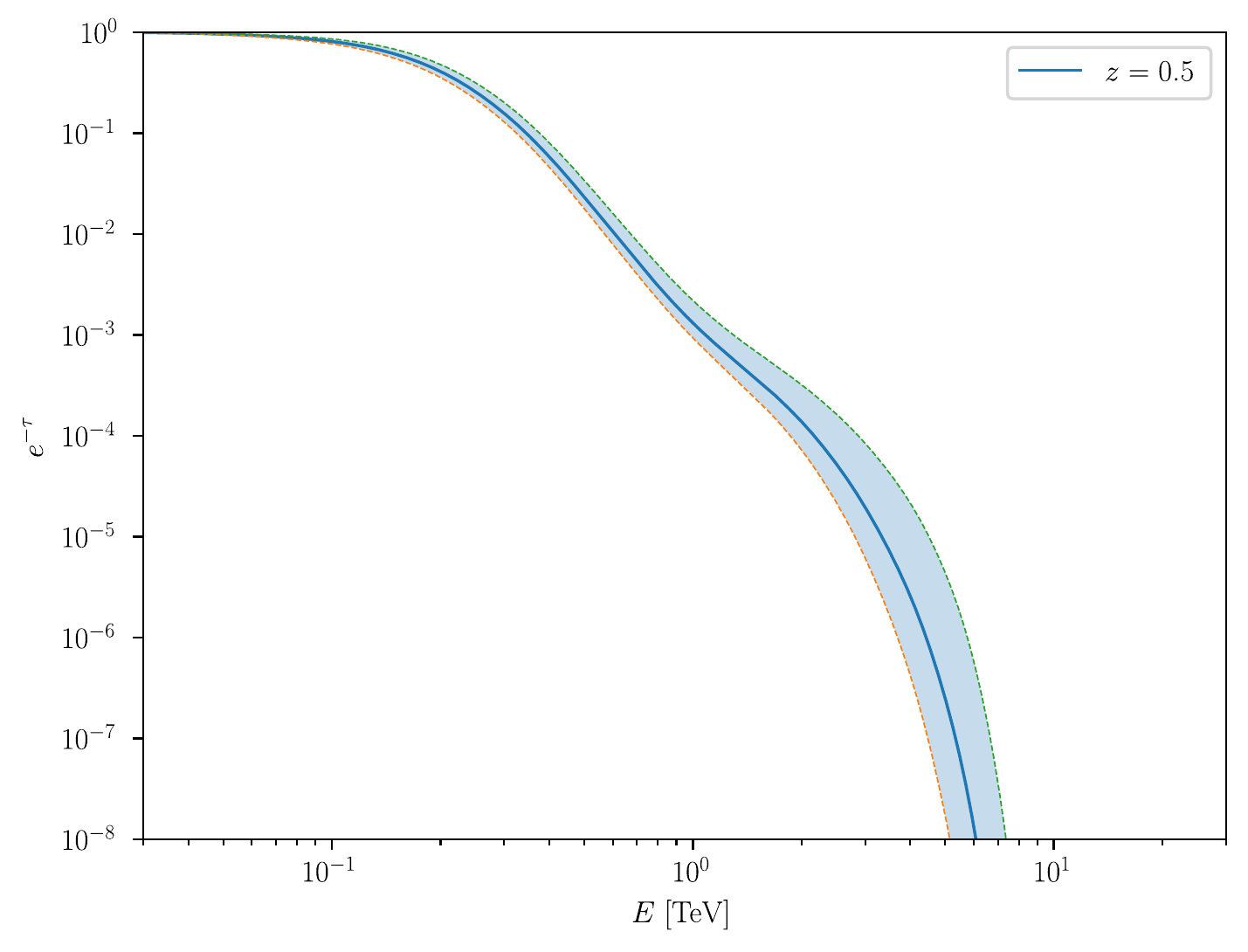}
  \caption{\label{tau0.5}The EBL attenuation~($e^{-\tau}$) of photons from a source with redshift $z=0.5$. The model and data are taken from Ref.~\cite{Dominguez2011} and \url{http://side.iaa.es/EBL/}.}
\end{figure}

\begin{figure}
  \includegraphics[scale=0.51]{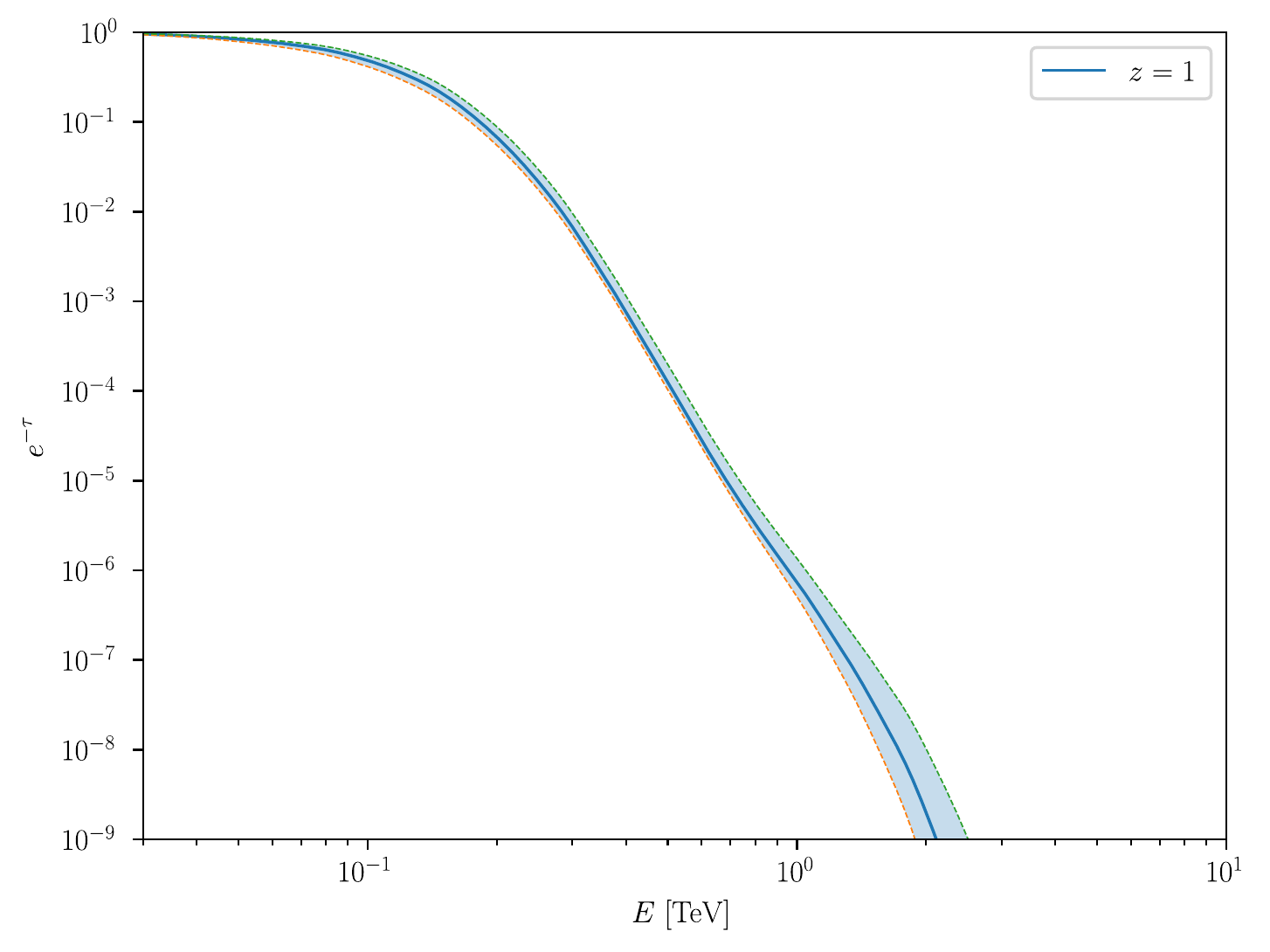}
  \caption{\label{tau1.0}The EBL attenuation~($e^{-\tau}$) of photons from a source with redshift $z=1.0$. The model and data are taken from Ref.~\cite{Dominguez2011} and \url{http://side.iaa.es/EBL/}.}
\end{figure}

\begin{figure}
  \includegraphics[scale=0.51]{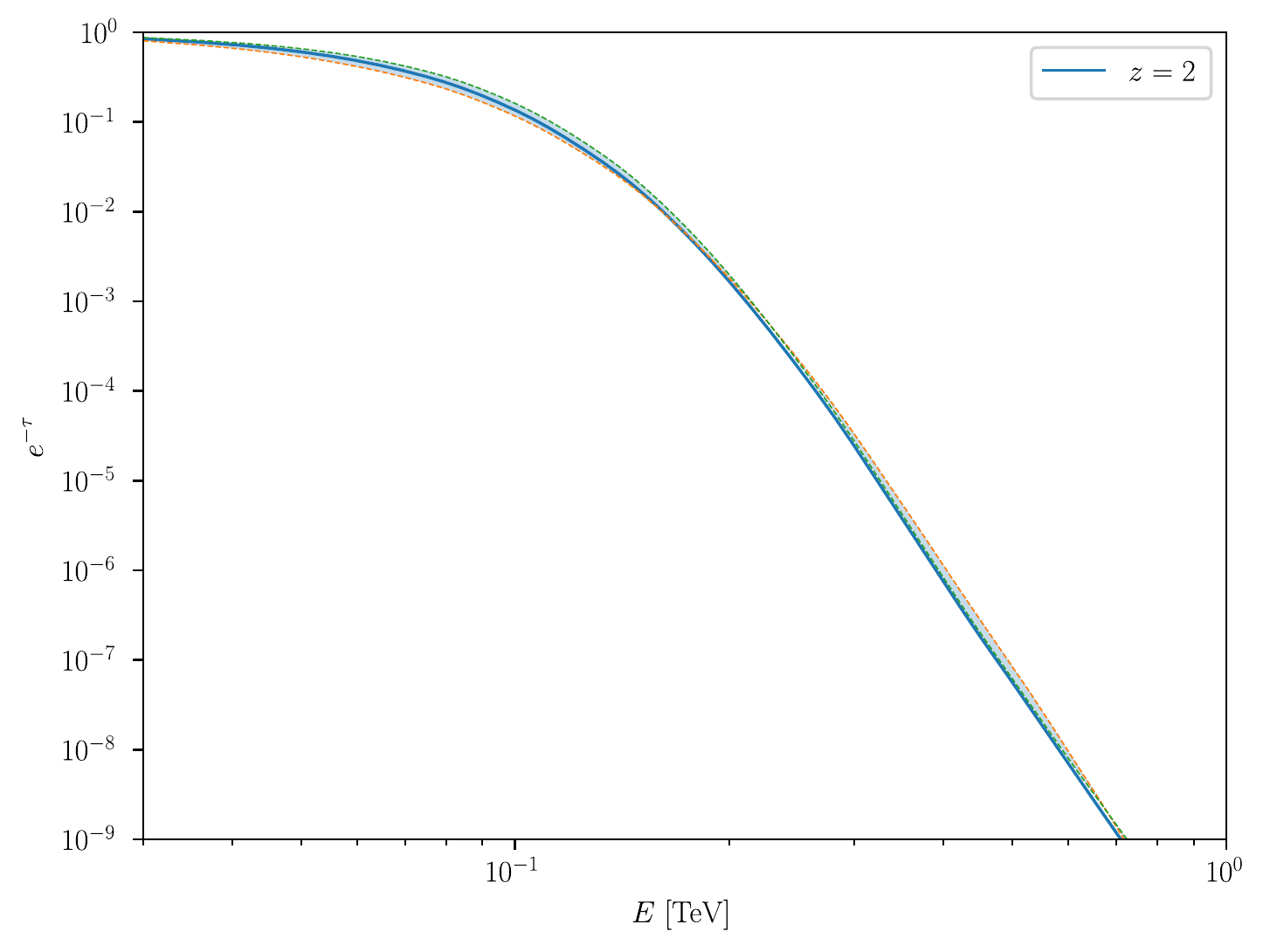}
  \caption{\label{tau2.0}The EBL attenuation~($e^{-\tau}$) of photons from a source with redshift $z=2.0$. The model and data are taken from Ref.~\cite{Dominguez2011} and \url{http://side.iaa.es/EBL/}.}
\end{figure}

\begin{figure*}
  \centering
  \includegraphics[scale=0.5]{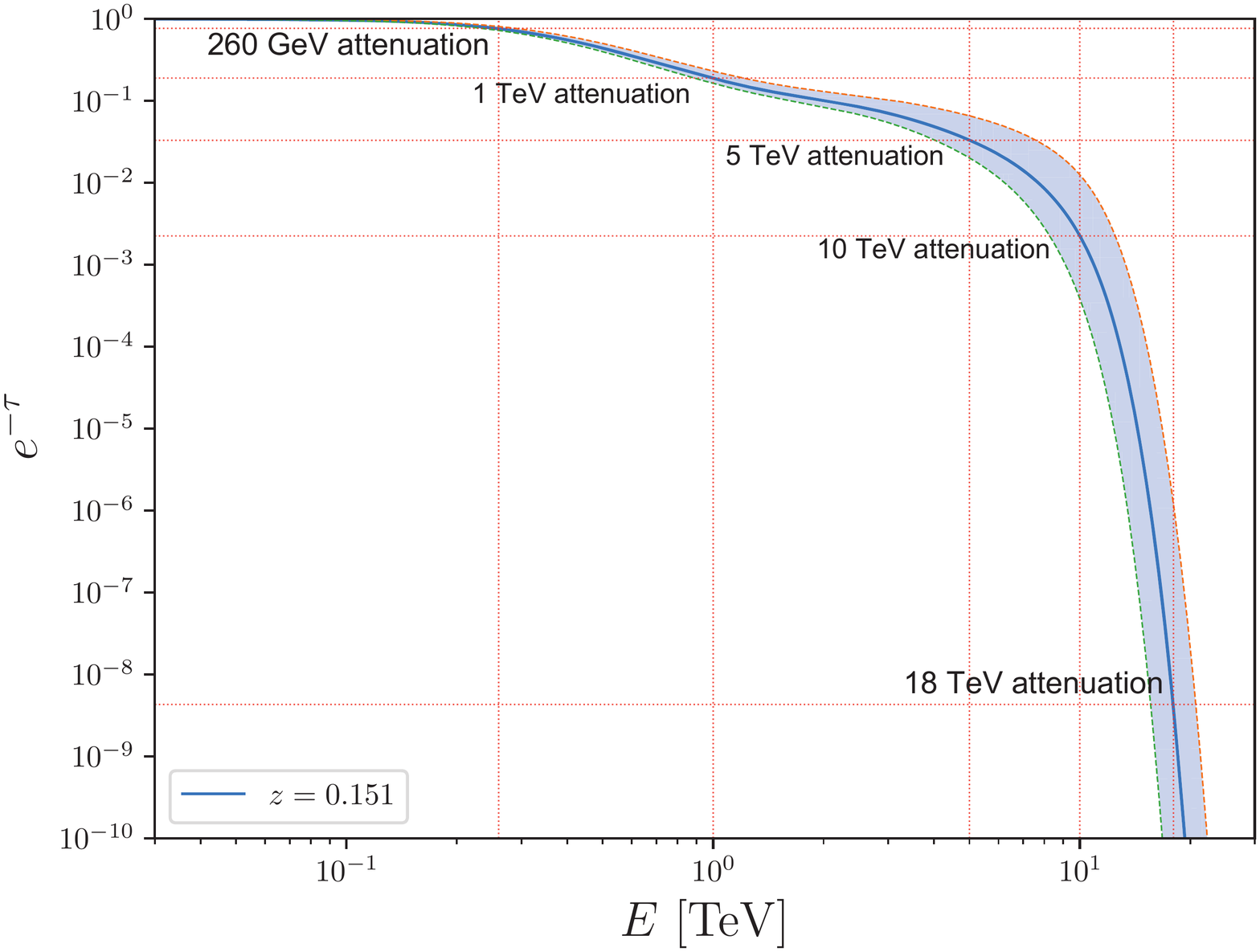}
  \caption{The EBL \(E\)-\(e^{-\tau}\) plot of redshift \(z=0.151\), with several representative positions marked out. The model and data are taken from Ref.~\cite{Dominguez2011} and \url{http://side.iaa.es/EBL/}.\label{tau0151}}
\end{figure*}

In order to search LIV from the data of LHAASO, we first utilize the strategies mentioned in \mysec{correlation} to correlate photon(s) to sources. Usually several sources may be correlated to one photon event or photons coming from the same direction and background analyses are needed. For convenience, we focus on the case that one photon is correlated to a source and the corresponding observed energy and redshift are $E_0$ and $z_0$ respectively. We consider if this event can be used to carry out the time of light flight study according to \myeq{tlv1} and \myeq{tlv3} just as in Refs.~\cite{Xu2016a, Xu2016} for example, but this situation may be not common since it requires the detailed knowledge of the source and generally if the source is a GRB with its redshift known, then we can accomplish this study. We then analyze whether it is reasonable that the photon can survive attenuations by EBL\@. Once any excess is revealed, threshold anomalies in \mysec{thresanomaly} can be considered as an explanation. Just as in both {\it Case I} and {\it Case II}, the background light can be more transparent to VHE photons than expected in SR and QED, and it is natural to detect excesses of VHE photons.
We show $e^{-\tau}$ for different redshifts in \myfig{tau0.04}--\myfig{tau2.0}. We take \new{$\exp (-\tau(E_0, z_0))=10^{-6}$} as a benchmark for excesses. If \new{$\exp (-\tau(E_0, z_0))=10^{-6}$} we conclude that we find potential excesses and further investigations such as spectrum analyses will follow. For typical redshifts, we can determine the benchmark energy $E_0$ using \myfig{tau0.04}--\myfig{tau2.0}. For $z_0=0.04$ in \myfig{tau0.04}, which is the case for AGNs such as Markarian 501 and Markarian 421, $E_0\simeq 40~\text{TeV}$; for $z_0=0.25$ in Fig.~\ref{tau0.25}, $E_0\simeq 11~\text{TeV}$; for $z_0=0.5$ in Fig.~\ref{tau0.5}, which is the mean value of short bursts, $E_0\simeq 4.5~\text{TeV}$; for $z_0=1$ in Fig.~\ref{tau1.0}, $E_0\simeq 1~\text{TeV}$; and for $z_0=2.0$ in Fig.~\ref{tau2.0}, which is \new{usually} the redshift of some long bursts, $E_0\simeq 0.4~\text{TeV}$. We can also find that $e^{-\tau}$ drops so fast when $E_0$ is greater than the corresponding benchmark that photons with higher energy can hardly survive for detection. In conclusion, if we can correlate photons with energy beyond the corresponding benchmarks to sources, we need to consider this as a signal of LIV and further analyses then should follow. It should be noted that this kind of study cannot be deterministic since the backgrounds are still remarkable, however, it can serve as a supplement to LIV phenomenology and provide circumstantial evidence.
\new{
  \subsection{Newly observed GRB 221009A\label{GRB221009A}}
  As a partial application of the strategy introduced in Sec.~\ref{correlation}, we analyze high energy photons from GRB 221009A~\cite{GBM1, GBM2, LAT1, LAT2, Swift1, Swift2, Swift3}, which is also observed by LHAASO~\cite{LHAASO2} at very high energies such that our method can be applied immediately. On 9 October 2022, GRB 221009A was reported by Fermi and Swift~~\cite{GBM1, GBM2, LAT1, LAT2, Swift1, Swift2, Swift3}, and observed by many other observatories, including LHAASO~\cite{LHAASO2}. This GRB, located at a redshift of about z=0.151~\cite{Redshift1}, is one of the nearest long bursts and is also one of the most energetic GRBs ever detected and its extraordinary brightness provides us with an unique window to study not only the mechanisms of GRBs themselves but also LIV induced threshold anomaly. The highest energy from GRB 221009A reported by Fermi reaches 99.3~GeV, while more surprisingly, LHAASO observes more than 5000 photons with energies above 500~GeV, including  a photon with energy up to 18~TeV, which is the most energetic GRB photon ever observed. The observation of such 18~TeV event raises a question naturally: do we need to invoke LIV induced threshold anomaly to explain the detection of a photon of such high energy? To address this question, we first consider the CMB attenuation of this photon. As is discussed, the corresponding threshold is around 411~TeV, which is obviously large enough so that a photon of 18~TeV can propagate freely in CMB. Then we consider the attenuation due to EBL which is also intensive and more energetic. For EBL photons, a high energy photon of about 261~GeV is enough to cause a pair production process and as a result the propagating distance of this photon is limited. Therefore the EBL attenuation of 18~TeV photons should be taken into consideration, and to discuss this situation numerically we separately exhibit the \(E\)-\(e^{-\tau}\) plot in Fig.~\ref{tau0151}. As is shown in Fig.~\ref{tau0151}, photons of 260~GeV from GRB 221009A is only suppressed by a factor of about 0.8, meaning that EBL is almost transparent to photons of energies around the threshold. Similarly the flux of 1 TeV photons is only suppressed several times and the flux of 5 TeV photons is merely suppressed dozens times, making these photons still easy to be observed. Indeed even for 10 TeV photons, suppression of several hundred times makes it still possible to be detected on the Earth. However the situation is completely different for 18~TeV photons: the flux of 18~TeV photons is suppressed by a factor of about \(10^{-8}\), and even consider the lower limit of EBL attenuation we still have a factor of about \(10^{-6}\). As a result, the EBL attenuation of 18~TeV photons obviously exceeds the benchmark we set, and it makes any observation of photons of or above 18~TeV from GRB 221009A a signal of new physics, a conclusion first pointed out in Ref.~\cite{Li:2022wxc}.
}

\new{
  There can be several theories to explain such 18~TeV photon from GRB 221009A~\cite{axion-cpl}, here we consider the LIV induced threshold anomaly introduced in Sec.~\ref{thresanomaly}~(see also~\cite[]{Li:2022wxc}). If the parameter \(\xi>\xi_c\),  
  then EBL is transparent to photons with energies of about 18~TeV, and consequently the observation of 18~TeV photons is just a result of LIV induced threshold anomaly. If the LIV effects are smaller, that is to say, \(\xi\) is smaller, thus we have {\it Case II}, i.e., LIV induced threshold anomaly can still explain 18~TeV photons since as long as \(\xi\) takes suitable values such that 18~TeV is larger than the upper threshold in {\it Case II}, EBL is again transparent to these 18~TeV photons. Of course it should be pointed out that, standard physics still has opportunities to explain the 18~TeV photons. For example, the EBL model we used should be modified or the 18~TeV photons are from near enough sources coincident with GRB 221009A both spatially and temporally or the actual energy of
  the observed event is around 10~TeV due to large uncertainty with energy reconstruction. 
}

\section{Conclusions\label{summary}}

In this work, we present a brief introduction to Lorentz invariance violation~(LIV) and LIV induced threshold anomalies, which can be an explanation for any excess of very-high energy~(VHE) photons observed by the Large High Altitude Air Shower Observatory~(LHAASO) in the future. We then give a review about VHE photon attenuations by extragalactic background light~(EBL) and a plausible model~\cite{Dominguez2011} and its results which serve as the fundamental material of our discussions. In \mysec{correlation}, we propose criteria for correlating VHE photons and potential sources after considering LIV which can avoid ignoring LIV effects if the standard methods are used. Based on these materials, we discuss how can we learn about LIV from the data of LHAASO\@. It is widely accepted that EBL attenuations can prevent photons from propagating from various sources; however, LIV induced threshold anomalies are possible to make this conclusion incorrect. As a result, it is hopeful that we can study LIV by searching excesses of photons that can hardly be detected for EBL attenuations. We find that TeV photons are sufficient for this purpose and LHAASO adjusts to this goal very well. Although we can only obtain circumstantial evidence from EBL attenuation studies, we expect that in the future along with the accumulation of data, we can utilize the criteria proposed in this work to provide more proofs, and by combining more methods, we can shed light on the study of LIV which eventually may provide guidance on the construction of the underlying theories.

Although in this work we only focus on LHAASO which is already in operation, we can also search LIV with the help of other observatories utilizing our strategies. Specifically, if an observatory can detect above TeV events and has the ability to distinguish photons from other particles, then our strategies can be adopted. These observatories include LHAASO, Cherenkov Telescope Array~(CTA)~\cite{Actis2011,ACHARYA20133,Fairbairn2014,CTAConsortium:2017dvg} and other future similar facilities to be constructed. Therefore we expect that our 
strategy
can help to understand LIV better with the help of more observatories, and abundant data in future will help us improve our criteria in turn. \new{Remarkably, in a new added section we apply the proposed strategy in this
  paper to discuss the newly observed GRB 221009A, demonstrating the
  predicted ability of revealing novel results from the observation of very high energy multi-TeV photon events by available facilities such as LHAASO. }

\backmatter

\bmhead{Acknowledgments}

This work is supported by National Nature Science Foundation of China~(Grant No.~12075003).

\bmhead{Data Availability Statement}

This manuscript has no associated data or the data will not be deposited. [Authors' comment: Sources of all the data present in this manuscript are provided and the original data can be easily found.]


\end{document}